# Very-High Dynamic Range, 10,000 frames/second Pixel Array Detector for Electron Microscopy

Hugh T. Philipp[1], Mark W. Tate[1], Katherine S. Shanks[1,6], Luigi Mele[2], Maurice Peemen[2], Pleun Dona[2], Reinout Hartong[2], Gerard van Veen[2], Yu-Tsun Shao[5], Zhen Chen[5], Julia Thom-Levy[3], David A. Muller[4,5], Sol M. Gruner[1,4,6]

[1.] Laboratory of Atomic and Solid State Physics (LASSP), Cornell University, Ithaca, NY, USA

[2.] R&D Laboratory, Thermo-Fisher Scientific, Achtseweg Noord 5, 5651GG Eindhoven, The Netherlands

[3.] Laboratory for Elementary-Particle Physics (LEPP), Cornell University, Ithaca, NY, USA

[4.] Kavli Institute at Cornell for Nanoscale Science, Ithaca, NY USA

[5.] School of Applied and Engineering Physics, Cornell University, Ithaca, NY, USA

[6.] Cornell High Energy Synchrotron Source (CHESS), Cornell University, Ithaca, NY, USA

**Abstract:**

*Precision and accuracy of quantitative scanning transmission electron microscopy (STEM) methods such as ptychography, and the mapping of electric, magnetic and strain fields depend on the dose. Reasonable acquisition time requires high beam current and the ability to quantitatively detect both large and minute changes in signal. A new hybrid pixel array detector (PAD), the second-generation Electron Microscope Pixel Array Detector (EMPAD-G2), addresses this challenge by advancing the technology of a previous generation PAD, the EMPAD. The EMPAD-G2 images continuously at a frame-rates up to 10 kHz with a dynamic range that spans from low-noise detection of single electrons to electron beam currents exceeding 180 pA per pixel, even at electron energies of 300 keV. The EMPAD-G2 enables rapid collection of high-quality STEM data that simultaneously contain full diffraction information from unsaturated bright field disks to usable Kikuchi bands and higher-order Laue zones. Test results from 80 to 300 keV are presented, as are first experimental results demonstrating ptychographic reconstructions, strain and polarization maps. We introduce a new information metric, the Maximum Usable Imaging Speed (MUIS), to identify when a detector becomes electron-starved, saturated or its pixel count is mismatched with the beam current.*

**Introduction:**

Hybrid pixel array detectors (PADs) have advanced scientific x-ray imaging at synchrotron light sources by offering low noise direct detection of photons coupled to custom signal processing electronics (Graafsma *et al.* 2020). Using this platform for electron imaging in scanning transmission electron microscopy (STEM) has enabled a major jump in data collection fidelity and speed (Jiang *et al.* 2018; Mir *et al.* 2016; Plotkin-Swing *et al.* 2020; Tate *et al.* 2016). At the heart of the technology is a hybrid PAD that uses a pixelated silicon sensor to directly absorb and detect incident energetic electrons with extremely low noise. The resulting electrical signal is collected and processed at the pixel level using customized CMOS electronics. The flexibility of analog and digital CMOS electronics offers many design choices and optimizations for different types of measurements. As a result, there are different types of PADs and detector performance depends on the specific design choices and optimizations that are reviewed elsewhere (Faruqi & Henderson 2007; Levin 2021).

One of the necessary design choices is how the pixel circuitry processes charge collected from the sensor, and two broad and fundamentally different approaches dominate PAD design. The first is counting of events, i.e., the detection of current pulses caused by discrete absorption of radiation quanta. This method relies on pulse shaping, thresholding of signal, and digital tallying of the total number of quanta detected. The second method is the integration of current in the pixel. This second method relies upon charge creation in the sensor that is proportional to absorbed energy. In integrating detectors, the pixel output is proportional to the total charge collected by the pixel. Both methods can have advantages and disadvantages, depending on the specifics of the experiment and what data are of interest. A key requirement for charge integration is that the sensor must be thick enough to collect all the deposited energy from the incident electron. If this condition is not met, the energy straggle follows a Landau distribution which for thin detectors becomes as large as the mean energy deposited (Bichsel 1988). The Landau distribution leads to large noise fluctuations that cannot be effectively suppressed even by summing multiple measurements. This problem is particularly noticeable in the analog output of thin monolithic active pixel sensor (MAPS) detectors (Bichsel 1988). At low count rates, this problem can be overcome by using pulse counters set to trigger if the deposited energy is above the thermal noise level. This is an effective strategy for low-dose imaging sensors, such as used in cryo-transmission electron microscopy. For the higher beam currents per pixel used in electron diffraction, electron energy loss spectroscopy and STEM imaging, pulse counting cannot reliably count all electrons that arrive at high rates, and the detector efficiency and noise performance degrade rapidly with increasing beam current. For high speed, or high beam current experiments, the integration of current by the pixel is favored because of difficulties of reliably counting quanta that arrive at high rates. Of course, for the charge integration strategy to work, the sensor must be thicker than the range of the electron, which at 300 keV is 452 µm in silicon. This is the strategy we have taken and describe in this paper.

The prototype imager described in this paper, the 2$^{nd}$ Generation Electron Microscope Pixel Array Detector (EMPAD-G2), uses current-integrating pixel circuitry and builds on the technology of an earlier generation EMPAD (Tate *et al.* 2016). This earlier generation EMPAD (1$^{st}$ generation) demonstrated collection and processing of 4D STEM data sets to provide center-of-mass (CoM), bright field, dark field, differential phase contrast, and full diffraction analysis. Applying advanced techniques like ptychography has yielded record-breaking microscopic

resolution(Jiang *et al.* 2018; Chen *et al.* 2021). EMPAD data has been analyzed for high resolution strain mapping over extended sample areas(Han *et al.* 2018) and to reconstruct magnetic and electric field distributions in samples(Nguyen *et al.* 2016a, 2016b). The first generation EMPAD was developed at Cornell and is available from Thermo Fisher. Like the EMPAD-G2 described in this paper, it is also a high-fidelity STEM imager. It is, however, limited to frame rates of 1.1 kHz and has a data-acquisition duty cycle that falls sharply above 1 kHz because the readout requires 860 µs to complete and the EMPAD is not designed to acquire new signal during readout.

The EMPAD-G2 prototype increases framing speeds to 10 kHz, extends the dynamic range, and allows for acquisition of signal during readout for a near-unity duty cycle even at 10 kHz. These capabilities allow for fast electron imaging of signals that vary by orders of magnitude across the face of the detector with almost no detector dead time. In practical terms, this allows for efficient high-speed, high-resolution raster imaging of extended areas. The speed of data acquisition and the dynamic range of the detector mitigates problems associated with sample stability by allowing high-quality, information-rich data to be collected quickly. The extension of critical performance metrics is expected to impact many types of STEM measurements. For many STEM applications, from ptychography(Chen *et al.* 2021) to strain(Padgett *et al.* 2020) and magnetic field(Xu *et al.* 2021) mapping, we find 128 x 128 pixels sufficient for high-resolution, high precision work. As noted previously for magnetic and strain mapping, and discussed in the section on the MUIS, the ability to deliver a high dose per pixel is more important than the number of pixels on the detector. Nevertheless, there are applications such as spectroscopy and continuous-rotation 3D electron diffraction where a larger pixel count is desirable. Our basic detector element has readout wiring along only one edge to allow for future stacking into tiled designs when larger pixel formats are needed. The EMPAD-G2 prototype increases framing speeds to 10 kHz, extends the dynamic range, and allows for acquisition of signal during readout for a near-unity duty cycle even at 10 kHz. These capabilities allow for fast electron imaging of signals that vary by orders of magnitude across the face of the detector with almost no detector dead time. In practical terms, this allows for efficient high-speed, high-resolution raster imaging of extended areas. The speed of data acquisition and the dynamic range of the detector mitigate problems associated with sample stability by allowing high-quality, information-rich data to be collected quickly. The extension of critical performance metrics is expected to impact on many types of STEM measurements.

This paper describes the design of the EMPAD-G2, the measured performance of the prototype, and examples of data acquired using the detector. These examples all demonstrate the need to work with higher beam currents when operating at higher speeds so as not leave the detector electron-starved. For atomic-resolution imaging, we want to record data as fast as possible to outrun environmental noise, but the faster we run the detector, the fewer electrons/pixel we will be able to record unless the counting or dose rate of the detector can be increased as well, as we have done so here. In mapping strains and fields, the ultimate precision depends on counting statistic and hence the dose delivered. Here, by increasing the maximum usable beam current on the detector, we show strain and polarization maps recorded at 100 $\mu$s/pixel instead of the more typical 10-100 ms needed to reach comparable precision. The resulting speed up reduces the acquisition time for typical 128x128 maps from 5-30 minutes down to under 2 seconds. This should be particularly valuable for in-situ experiments, or when timing is less critical, more detailed maps can be recorded in the same time.

We also introduce a measure that describes the rate at which the detector can collect information – the maximum usable imaging speed (MUIS) at which the detector can reach a desired signal-to-noise ratio (SNR). We have found this helpful in thinking about detector design strategies and addressing questions such as how many pixels can be usefully illuminated. Usually, detector performance as a function of dose is described in terms of dynamic range, but this gives no indication of how long it will take to deliver sufficient electrons to fill the dynamic range. This can sometimes be as long as 20-30 seconds dwell time per frame, which is a far cry from the millisecond operating times expected for 4D-STEM mapping. Reporting the saturation current per pixel can be helpful to ameliorate this problem and should be done. However, when there is a soft roll-off in linearity, as, for instance, with pulse counting detectors, there can be an order-of-magnitude difference in where to define the saturation level. The ambiguity can be resolved by properly accounting for the loss of detective quantum efficiency when the output signal becomes sublinear. The MUIS can capture these details, making it simple to trade-off pixel count for SNR when the detector is electron starved, or increasing the pixel count if individual pixels are saturating, with an end goal of reaching the desired SNR in shortest possible time. The EMPAD-G2 retains a high MUIS across a wide range of SNRs, allowing very high precision field measurements to be performed at speeds more typically associated with imaging (0.1 ms per pixel) than traditional quantitative mapping (10-100 ms per pixel).

**Materials and Methods:**

**Detector Description:**

The EMPAD-G2, like all hybrid PADs, comprises two functional layers. The first layer is a sensor layer that absorbs incident radiation, converting the absorbed energy to electron-hole pairs. The second layer is a custom CMOS integrated circuit (IC) that collects the charge generated in the sensor-layer and converts it into readable information that can be used to construct quantitative images. To ensure complete energy transfer and minimum energy straggle, the sensor layer is chosen to be thicker than the 452-µm range of a 300 keV electron. This can be done without compromising the lateral point spread function compared to the typical PAD sensor thickness of 350 µm as the incident beam's maximum spread occurs at about half the range. The sensor layer is a 500 µm-thick, high-resistivity, silicon diode that is pixelated on one side. The pixelated side mates to the signal processing CMOS, which is also pixelated. The pixel size is $150 \times 150$ microns. In operation, the silicon diode (i.e., sensor) is kept fully depleted by reverse biasing the diode with high voltage applied to the detector face. Typical reverse bias voltages are between 150 and 200 V. The sensor is fabricated to specification by SINTEF (Trondheim, Norway).

The CMOS Application Specific Integrated Circuit (ASIC) layer of the electronics is fabricated by Taiwan Semiconductor Manufacturing Corporation (TSMC) using a 0.18 micron mixed-mode process. The full monolithic CMOS die has 128 × 128 pixels, matching the pixel-by-pixel format of the Si sensor.

The sensor and the CMOS layers are mated to one another, pixel-by-pixel, using an array of solder bump bonds. The bumps are lithographically fabricated on the fully fabricated TSMC CMOS wafer by Micross Advanced Interconnect Technology LLC (Research Triangle Park, NC). Micross also processes the sensor wafer to apply a pixel-level metallization that is compatible with the bumps on the CMOS wafer. After processing, the wafers are singulated to make compatible CMOS and sensor dies. The dies are mated using a flip-chip process, and the resulting hybrid detector module is mounted on a heatsink and wire bonded to a printed circuit board that conveys the signals necessary for operating the chip and reading data. Signals supplied to the ASIC include voltage and current biases for analog and digital components; and digital waveforms for chip operation that allow for the synchronization of image acquisition with external systems (e.g., electron microscope scanning). Output signals from the ASIC include 16 differential analog and 16 LVDS digital data outputs; and two additional LVDS clocking outputs for synchronizing the 200 MHz digital data from the LVDS data outputs.

The detector module is actively cooled by a miniature thermoelectric cooler. The cooler is attached underneath the module and held to -20 ± 0.1 C via a tuned thermal feedback loop. An external chilled water circulator is used to remove heat from the thermoelectric unit. The detector module assembly is attached to a pneumatic actuator that allows for in-vacuum insertion into the microscope or retraction into a radiation shielded shroud.

**Pixel operation:**

The design of the EMPAD-G2 CMOS pixel offers several advances over the previous EMPAD, including a higher frame rate that reduces scan time; extended dynamic range that allows use of higher EM beam currents; and the ability to acquire data during readout, greatly reducing detector deadtime and speeding up STEM data set collection. This is important because many applications require sample stability at the sub-Angstrom level over the data set collection time, thus depending on rapid data acquisition. The dynamic range metric that is relevant to these types of high-speed measurements is defined by incident power on the detector, not simply a statement of well depth or number of bits in a digital counter.

The high-level pixel diagram shown in Figure 2 indicates how some new detector capabilities are accomplished. The easiest way to describe pixel operation is by tracking the processing of collected charge through the schematic. The charge enters the pixel through a bump bond and is collected by an analog integrator that has one capacitor (40 fF) actively in feedback loop and another that is primed to be switched into the feedback circuit. Both capacitors are cleared of charge before acquisition. If the output of the integrator passes a threshold voltage, $V_{th}$, during acquisition, the second capacitor (840 fF) is switched into the feedback of the front-end integrator, lowering the gain of the front-end integrator and extending the dynamic range of the analog front-end. This scheme is similar to that used by the x-ray adaptive gain integrating pixel detector (AGIPD)(Trunk *et al.* 2017). If the front-end is in low gain and the Vth is passed again, a switched capacitor charge dump circuit is triggered that extracts a bolus of charge from the front-end without breaking the feedback loop of the integration stage, so that integration continues uninterrupted. Every subsequent passing of Vth also triggers the charge dump circuit. Each time a charge dump occurs, an in-pixel counter is incremented. Charge dumping can happen at rates up to $10^8$ dumps per second, a hundred times faster than in the original EMPAD, resulting in a significant extension of the dynamic range of incident electron current. The output of the pixel is the combination of the remaining signal in the integrator at the end of the frame (conveyed as a residual voltage from the pixel differentially referenced to a reference voltage), a digital gain bit (that conveys what gain state the pixel is in), and a 16-bit word that encodes how many charge dumps have happened during acquisition. These data are merged using calibration constants to yield a smooth, linear, monotonic signal proportional to the incident electron energy deposited in the sensor.

In addition to the basic signal processing, additional features allow for acquisition of signal during readout. First, two 16-bit counters are alternately used in successive frames so that the value of one is being read out while the other is actively counting charge dumps of the next image. Second, the analog voltage is sampled onto one of two in-pixel track-and-hold circuits at the end of any given frame. While one of the track-and-hold circuits is tracking the output of the front-end integrating stage, the other is holding the value of the previous frame and is read out. This in-pixel double buffering of both digital and analog values allows for very high duty cycle active detection of more up to 99%, while framing at 10,000 frames per second. The gain bit is latched into one of two additional pixel status bits and shifted out with the counter data. As a result, the pixel produces 18 bits of digital data. Since the analog data is digitized to 14-bits using a pipelined off-chip ADC, each pixel yields a 32-bit data value for each frame.

**Readout structure:**

Readout of the CMOS ASIC requires both analog and digital readout. These are performed in parallel and independently, meaning that waveforms for each readout have no predetermined phase with respect to one another, other than that imposed by the acquisition of frames. The ASIC is composed of $128 \times 128$ pixels organized into 16 separate banks with $8 \times 128$ pixels each. The banks are read out in parallel with one digital LVDS pair and one analog differential pair for each bank.

The analog readout structure, shown in Figure 3, consists of the dual track-and-hold circuits (discussed in the pixel description above), a dual track-and-hold multiplexer at the top of a bank, and a differential output amplifier. The in-pixel dual track-and-hold circuits alternate with each frame, so that during the readout of a single frame the selection is static, i.e., a single analog value is ready to be read. Addressing of pixels to be read out is done by a row-select signal that is fanned out to all pixels in the row, with a row defined as the shorter dimension in an $8 \times 128$ pixel bank. The addressing of the column is accomplished with the dual $8 \times 1$ mux. The reason for dual track and hold at the column level is to allow pre-charging of the analog lines before sampling while previously sampled values are being read out. This mitigates the effects of parasitic time constants and produces a clean sample-and-hold signal that is fed into a differential amplifier. Both the signal from the pixel and the reference voltage from the pixel are sampled in parallel. Analog values are converted to digital values off-chip at a 10 MHz rate.

The digital readout scheme, diagrammed in Figure 4, consists of two in-pixel counters that are readout as shift registers on alternate frames. Each of these, arbitrarily designated as data streams A and B, are daisy-chained with all pixels in the same column during readout. On the edge of the chip, a pixel worth of bits (i.e., 18 bits) are shifted into a shift register at the column edge. This shift register is then daisy-chained with all similar shift registers in the bank and read out at approximately 200 MHz. While these bits are shifted out, a clock running at approximately $1/8^{th}$ the speed shifts data from the array into another set of shift-registers. These shift registers follow the same process and are multiplexed into a daisy-chain for 200 MHz readout. This scheme has the advantages of feeding a slower clock signal (200 MHz/ 8 = 25 MHz) through the array, and keeping bits associated with a particular pixel grouped together. This second advantage reduces the need to re-order bits in the readout FPGA and simplifies trouble shooting, if needed, because pixel outputs are easily isolated on an oscilloscope.

**Support electronics:**

The wire bonds of the ASIC that connect to the signal processing electronics are all on the same side of the detector module, allowing for potential 3-side tiling of modules. All power, biases, digital control, and signal outputs from the chip are wire-bonded from this single edge of the module to a PC board that has appropriate buffers, digital-to-analog converters (DACs), and power biases for the chip. This board resides in the vacuum and connects to a feedthrough board that provides electrical connections through a vacuum flange to the PC board with analog to digital converters, voltage regulators, and an FPGA that manages the low-level waveform operation. A fiber optical link from this board provides a GenICam, Generic Interface for Cameras standard("GenICam – EMVA" n.d.), compliant control interface for the detector. Data is captured using a frame grabber board.

**Results and Discussion:**

**Data combination and calibration:**

Raw data from the array comprises the analog signal from the amplifier, the gain bit, and the output from the digital counter which indicates the number of times a charge removal operation was performed (a 16-bit word). To calibrate the scaling factors needed to combine the raw data into a linear response, a data set is taken with constant illumination and increasing integration time (Figure 5). The data shown were obtained using an optical LED array (Bridgelux, BXRC-50C1001-D-74-SE), run at a constant current. This optical flood field is not completely uniform because of variations in the entrance window metallization, but uniformity in this calibration step is not needed. A source which is stable in time is required. Optical photons have the added benefit of providing a signal with much less Poisson shot noise than an equivalent illumination with high-energy electrons, reducing the number of frames needed to average to obtain the scaling factors to high precision.

Linear regression is applied to the three different signal domains: high-gain analog (analog when the gain bit equals 0), low-gain analog (analog when the gain bit equals 1), and digital. With these regressions, all signals are scaled to equivalent high gain analog-to-digital units to produce a continuous linear output (Figure 5D). Each pixel has unique scaling coefficients arising from fabrication process variations (e.g., variations in capacitor sizes). Also, double buffering leads to two unique sets of analog sampling circuitry, so each pixel requires two sets of calibration coefficients.

The above procedure produces a linear output for each pixel, scaled to the output voltage of that pixel. The relative gain between pixels can vary, so a final scaling factor is needed to normalize all pixels to the same scale. This calibration can be obtained using histograms of the response of each pixel to single electrons, as shown in the next section. The position of the single electron peak is directly proportional to the absolute gain of each pixel. It was found in practice that the pixel normalization coefficients could be determined to higher precision using the optical flat field illumination rather than a defocused source of electrons (3% precision for coefficients determined by electron histograms vs $< 0.1\%$ precision by using flood field calibration).

**Low-fluence electron microscopy measurements:**

Figure 6 shows low fluence measurements made with a wide aperture in the Thermo Fisher Themis CryoS/TEM electron microscope. The aperture was chosen such that a roughly uniform illumination was incident on the detector. Signal levels less than 0.1 electrons/pixel/frame on average are needed to avoid substantial overlap of individual electron events. Three electron energies were used: 300, 120, and 80 keV. Figure 6A shows histograms of single pixel outputs gathered from the full array over 50,000 frames. Background pedestal subtraction was applied to this data set, with the pedestal measured by taking frames with no incident electrons present (i.e., a dark frame). Pixel calibrations described in the previous sections were also applied. There is a zero-electron peak on the left corresponding to no detected charge from incident radiation and, for the different energies, an integer number of peaks to the right. The positioning of the peaks is determined by the energy of the detected electrons, meaning 300 keV electrons deposit proportionately more energy and produce a proportionately higher signal than 80 or 120 keV electrons.

The distinctness of the peaks is also a function of energy because the charge produced in the silicon sensor can spread over adjacent pixels. The area over which charge is likely to spread increases with electron energy and only a few events at 300 keV will be contained within a single pixel. The histograms are a mapping of these stochastic processes projected onto many thousands of measurements. In other words, the charge resulting from single incident electrons can spread over multiple pixels. For energetic electrons, this spread depends primarily on a random walk the incident electron takes through the silicon as it loses energy and produces collectable charge. Over a large number of frames, the sum of these random walks can be viewed as producing a probabilistic distribution of deposited charge. Figure 6C shows single-electron events at 300 keV for a single frame and a small sub-section of the imaging area.

A cluster analysis can be performed on these images to provide a histogram of the total energy deposited per electron event (Figure 6B). Individual events are detected, a local area around each is defined and the signal from each event is summed using the OpenCV(Culjak *et al.* 2012) connected components algorithms(Bolelli *et al.* 2017). This recombines the charge deposited from single electrons that has been split between pixels. The single electron peak is much more distinct and symmetric than in Figure 6A. In this plot, electrons which fully deposit their energy within the sensor contribute to the peak, whereas electrons which lose energy due to other processes (e.g., florescence or back-scatter) contribute to the low energy tail. The peak position is found to be 3661 ADU for 300 keV, 1453 for 120 keV and 960 for 80 keV. Using the shape of the tail in the distributions, the mean signal per recorded electron is 3262 ADU for 300 keV, 1258 ADU for 120 keV, and 832 ADU for 80 keV.

**Linearity of response at high flux:**

To measure the linearity of response to increasing beam current, a small (< 4 pixels FWHM) focused spot of 300 keV electrons was imaged at beam current settings that varied over three orders of magnitude. A cross section of the spot is shown in Figure 7 (left). As the beam intensity increases, the signal within a pixel continues to increase up to a maximum rate determined by the speed of the charge dump circuitry in each pixel. At beam currents above this rate, the pixel response will saturate. An independent measure of the beam current was obtained by recording the current flowing from the sensor power supply (Keithley 2400 source meter). This supply shows a linear response well beyond the limit set by the pixel circuitry. The sensor current will have a gain of 8.33 x $10^4$ with respect to the beam current since an electron-hole pair is created in the sensor for every 3.6 eV of incident electron energy. At each beam current, a set of 100 µs exposures were averaged over 1000 frames. The intensity in the brightest pixel was converted to a primary electron current over this time using the gain factor for 1 electron obtained from single-electron histograms shown in Figure 6. Figure 7 (right) shows the current in the brightest pixel as a function of total sensor current. Response is linear up to 175 pA/pixel of 300 keV incident electron beam current, at which point the response saturates.

Pixel saturation is a function of biasing levels supplied to the signal processing electronics of the ASIC. These measurements were made at nominal settings and it should be noted that adjusting biases can affect (both increase and decrease) the saturation level. In-pixel bias settings do affect other properties (e.g., uniformity and pixel gain) with these nominal settings chosen for good overall performance. All characterizations in this paper were taken with the same bias settings. The maximum usable primary beam current also scales inversely with the incident electron beam's energy, so at 60 keV the saturation beam current would be around 875 pA/pixel. Written explicitly, the integrated signal incident on a single pixel at the maximum measurable beam current and full frame rate is:

$$S_{max} = \left(175 \times 10^{-12} \frac{coulomb}{s\ pixel}\right)\left(\frac{10^{-4}\ s}{frame}\right)\left(\frac{1\ e^-}{1.6 \times 10^{-19}\ coulomb}\right)\left(\frac{300\ keV}{1\ e^-}\right)$$

$$= 3.3 \times 10^7\ keV/pixel/frame$$

Detecting individual electrons allows for high fidelity measurements, but the real strength of a high dynamic range detector is combining low fluence (i.e., single electron) detection with the ability to quantify high intensity signals in the same frame. Looking again at Figure 7 (left), we see the profile of the spot is measured over 6 orders of magnitude. The tails show a fairly uniform floor at < 0.03 electrons/pixel (i.e., an electron strikes a pixel in this region only once in every thirty images on average). Importantly, the dynamic range shown in Figure 7 is realizable at a 10 kHz frame rate (100 µs frame time). As noted in Table 1, the dynamic range of the pixels at a 10 kHz frame rate is $1.3 \times 10^7$, calculated by taking the ratio of the highest measurable signal (175 pA incident current) and noise of a detector pixel in equivalent keV, 2.6 keV.

**Spatial resolution:**

The spatial resolution of the EMPAD-G2 is a function of both the pixel size and the spread of charge when incident electrons interact with the 500 µm thick silicon sensor. Each incident electron undergoes a random walk through the sensor. When taken as an ensemble, the r.m.s. width of the charge spread increases with increasing incident electron energy. The spatial resolution was measured by imaging a sharp-edged, nominally circular aperture at three energies (80, 120 and 300 keV, Figure 8A). The aperture edge was fit to a circle and the one-dimensional (1-D) edge-spread function (ESF) was found by plotting the intensity of a pixel in the image vs the distance of that pixel from the fit circle (Figure 8B). This method allows the edge spread function to be sampled much more finely than the size of the pixel. The edge spread function was fit to the convolution of a linear ramp (ramping from 1 to 0 over the width of one pixel) and a Gaussian function. The widths of the Gaussian function in these fits show an r.m.s. charge spread of 201 µm, 67 µm, and 44 µm for 300 keV, 120 keV, and 80 keV, respectively.

The line spread function (LSF) can be computed by differentiating the ESF. Here we differentiate the fitted function as a method to smooth the sampling noise of the data (Figure 8C). For high dynamic range imaging, the low-level tails to the LSF are important quantities as they determine how far a weak signal must be from a strong signal before it can be seen. One can measure the full width at 1/100 maximum (FWCM) and the full width at 1/1000 maximum (FWKM). For 300 keV electrons, the FWCM is 4.4 pixels and the FWKM is 5.6 pixels. These are reduced to 1.8 and 2.1 pixels for 120 keV and 1.6 and 1.8 pixels for 80 keV.

The modulation transfer function (MTF) was computed by taking the Fourier transform of the LSF (figure 8D).

**Detective Quantum Efficiency:**

The precision of any measurement is Poisson limited by the number of primary quanta in the signal. For M incident electrons, the shot noise scales as sqrt(M). How well a detector achieves this ideal performance is quantified by measuring the detective quantum efficiency (DQE), defined by

$$DQE = (S/N_{output})^2 / (S/N_{input})^2 \qquad 1$$

where $S/N_{output}$ is the signal to noise ratio as recorded by the detector, and $S/N_{input}$ is signal to noise ratio of the incident signal. With a Poisson distribution for the incident electrons, this reduces to

$$DQE = (S/N_{output})^2 / M \qquad 2$$

In general, DQE will be a function of electron energy, spatial frequency and total dose recorded. DQE as a function of spatial frequency, ω, is usually computed by

$$DQE(\omega) = DQE(0) \times MTF(\omega)^2 / NPS(\omega) \qquad 3$$

where DQE(0) is the DQE at zero spatial frequency, NPS(ω) is the normalized noise power spectrum and MTF(ω) is the modulation transfer function.

The noise power spectrum was calculated by taking the 2d-FFT of the difference of two nominally uniform illuminations. This was averaged from the FFTs of 200 to 5000 difference pairs at each energy. The 1-d NPS was found by taking the azimuthal average of the 2d FFT.

DQE(0) is found using equation 1 above using the uniform illumination data set. DQE is calculated with regions of interest spanning $1 \times 1$ pixels to $50 \times 50$ pixels. DQE(0) is taken as the asymptotic value found as the regions size become larger. Images are taken pairwise, with the signal found from the sum of the pair, and the noise computed from the difference. Taking the difference will eliminate systematic variations within the flood illumination. The incident signal in each region of interest is found from the average signal in each region, normalized by the average signal per incident electron found from single-electron event histograms. DQE(0) is found to be 0.94 for 300 keV, 0.9 for 120 keV and 0.9 for 80 keV. DQE($\omega$) is shown in figure 9 for each of these energies.

**Maximum Usable Imaging Speed:**

An important criterion in designing and operating a detector is how many electrons we can deliver to a given pixel in a given frame exposure time—if too many electrons arrive in the given interval then the detector will saturate, and if too few electrons are delivered, we are wasting readout bandwidth, storage memory and risking adding additional and unnecessary noise. Dynamic range, when defined as the ratio of largest to smallest detectable signal in a frame with indeterminate frame rate, is not sufficient to capture this effect—for instance, if a counting detector saturates at a count rate of 1 MHz but has a dynamic range of 24 bits then it will take over 16 seconds to fill the dynamic range. We also need a detector that can tolerate a high beam current so reasonable frame rate can be achieved. Here we introduce a metric that captures both of these requirements, which can be helpful for matching source and detector to reach the desired information quality needed for a particular experiment. This is the maximum usable imaging speed (MUIS) at which a particular signal to noise ratio can be reached.

The S/N$_{output}$ will depend on the number of electrons collected and the DQE of the detector. For pulse-counting detectors, the DQE is often quoted for the very-low-fluence limit because it degrades as a function of fluence (Li *et al.* 2013). Equation 2 can be modified to capture this trend by noting that if some counts are missed and the dead times are uncorrelated, then if only a fraction $\eta$ of electrons are counted then $S/N_{output}(\eta) = (\eta M)/\sqrt{\eta M} = \sqrt{\eta M}$ instead of $\sqrt{M}$. Substituting into equation 2, the DQE at a collection efficiency $\eta$ is related to the DQE at low fluence

$$DQE(\eta) = \left(S/N_{output}(\eta)\right)^2 / M = \eta \, DQE_{low} \qquad 4$$

This expression holds for low dead times, i.e. high collection efficiencies ($\eta \gtrsim 0.7$), but once the signal becomes noticeably non-linear the DQE degrades exponentially, reflecting the exponential sensitivity to noise in attempting to correct non-linearities(Li *et al.* 2013).

The question of what is the maximum speed we can operate at to reach a desired $S/N_{output}$ now becomes the question of what is the maximum speed at which M e$^-$ can be delivered to the pixel? That is to reach a signal/noise ratio of SNR, the number of incident electrons needed is

$$M = \frac{SNR^2}{\eta \, DQE_{low}} \qquad 5$$

The shortest frame time in which M electrons/pixel can be captured gives the MUIS:

$$MUIS(SNR) = \left(\frac{I}{e^-}\right) \frac{\eta \, DQE_{low}}{SNR^2} \qquad 6$$

where *I* is the incident beam current/pixel in Amps. For instance, the Rose criterion for imaging requires a SNR=5. If we had an ideal detector that operated at 100 kHz but only counted at most 1 electron/pixel/frame for a saturation current of 16 fA/pixel, it would take 25 frames to reach Rose criterion, and the Rose speed or *MUIS(SNR=5)* would be only 4 kHz.

One strategy to increase the MUIS is to reduce the pixel count, though potentially at the sacrifice of momentum resolution. For a total beam current, $I_{tot}$, and *n x n* pixels in the detector, the beam current per pixel can be written as $I = I_{tot}/n^2$. Substituting into equation 6, we see that reducing *n* increases the MUIS quadratically, so long as each individual pixel can handle the increased beam current without saturation. To put it another way, doubling the number of pixels in each direction will reduce the MUIS by a factor of 4 if the incident beam current is unchanged.

The EMPAD-G2, operating at 10 kHz frame rate, has a saturation current/pixel of 175 pA at 300 keV allowing for a signal to noise ratio per pixel of over 300. i.e. MUIS(SNR=300) would be 10 kHz. Only when a signal to noise ratio/pixel greater than 300 was needed would the imaging speed drop below 10 kHz. For instance, if we were trying to resolve the diffuse scattering in a diffraction pattern simultaneously with the details of the central disk, we might require a SNR of 1,000, and the MUIS(SNR=1000) would now be a little over 1 kHz as shown in Figure 10a.

In Figure 10 we explore the MUIS attainable for different detector strategies, including the EMPAD and EMPAD-G2. We also consider:

- a state-of-the-art pulse counting detector operating with 8-bit collection for high-speed sampling and with DQE=0.8 and $\eta$=0.55 at 1 pA input current/pixel. i.e. at 1 pA, 55% of incident electrons are counted. We re-bin over 16 pixels to compare to the 128x128 EMPAD. This is labelled "8-bit pulse" in Figure 10,
- a MAPS detector pulse counting at 1 e$^-$/pixel/frame sampled at 87 kHz, and re-binned by 16 as well. This is labelled "1-bit MAPS" in Figure 10
- a large-pixel format MAPS detector, such as that typically used in cryo electron microscopy with a maximum count rate of 30 e$^-$/pixel/frame, a frame rate of 1.5 kHz and re-binned by 256. This is labelled "MAPS" in Figure 10.

There are many more permutations of designs to consider. For instance, if the readout speed is limited by the data transfer bandwidth, doubling the bit-depth of the signal and halving the frame rate can lead to a significantly larger MUIS at large SNR. To capture different design choices on a single plot, we summarize the performance of each combination by its MUIS at SNR=5 vs SNR=300 in Figure 10b. This reflects two extreme limits of possible use cases—SNR=5 for high-speed but noisy readout for TEM imaging or simple STEM imaging modes like atomic-resolution, CoM where the signal will be integrated over the detector plane, and SNR=300 for quantitative measurements of strain and magnetic fields where high doses are needed for high precision. For quantitative work, the EMPAD-G2 can reach the needed SNR roughly two orders of magnitude faster than the other designs.

Again, it is worth noting that when the SNR per pixel drops well below the Rose criterion of SNR=5, then the detector is too electron starved to make effective use of such small pixels, and either a larger pixel size should be chosen, or a sparser readout scheme employed to increase the frame rate. Currently detector frame rates are limited by the data transfer bandwidth, so the fastest detectors are currently quadrant detectors, i.e., $2 \times 2$ pixels, with discrete readout electronics, and these can reach readout speeds of ~20 MHz with a nanoamp of beam current. This would give a *MUIS$_{quad}$(SNR=5)* of 20 MHz, and a *MUIS$_{quad}$(SNR=300)* of 55 kHz. Given that the differential phase contrast (DPC) output of the quadrant detector is visually almost indistinguishable from the CoM analysis from a pixelated detector at low to moderate signals, *MUIS$_{quad}$* serves as a useful guideline as when to use a quadrant detector, and when to use a multi-pixel direct detector. For present detector technologies, using a quadrant detector for high-speed, low-dose DPC imaging outperforms a PAD using CoM (as the PAD *MUIS(SNR=5)* would only reach 20 kHz), especially since live frame averaging and data storage becomes much simpler to manage. However low-dose, widefield ptychography where the large-pixel-number format is exploited to avoid sampling at every spatial point will outperform the quadrant detector in required dose and collection speed(Chen *et al.* 2020).

The source performance can also be captured on this type of plot for different imaging modes. SNR=5 can be used to evaluate the maximum frame rate that can be delivered to a uniformly illuminated detector with 128 x 128 pixels, while SNR=300 is useful for diffraction experiments when the incident beam is focused into a few pixels—here assumed to be 4 pixels. The expected performance limits for a cold field emission source are shown as the bounds for Figure 10b. From this we can see that there is considerable room for improvement in detector technology, both in frame rate and saturation current, before performance becomes source limited. Pixel count can be traded for speed, provided the necessary current/pixel can be maintained as discussed above.

**Experimental data**:

As a demonstration of the detector sensitivity and dynamic range, Figure 10 shows the convergent beam electron diffraction (CEBD) patterns of $[101]_O$ TbScO$_3$ recorded with 300 keV electrons and a beam current of 1 nA so high-quality patterns can be recorded with a short dwell time, taking advantage of the good MUIS metric for the detector. Figures 11a & 11b show the CBED pattern recorded in 100 µs and displayed in logarithmic scale, and in units of number of electrons. Even at 100 µs, the CBED pattern shows both the unsaturated central beam and intensity variations in the Bragg reflections (Figure 11a), as well as the details of Kikuchi bands and high-order Laue zone (HOLZ) rings, while still retaining an unsaturated central beam (Figure 11b). In particular, Figure 11b shows an unsaturated, undistorted central peak with 50 x 10$^6$ e$^-$/s/pixel, well beyond the possible linear or correctable count rate for a pulse-counting detector. In addition, it is not even close to the saturation limit for the EMPAD-G2, which is $10^9$ e$^-$/s/pixel at 300 kV – we would get closer to this limit for the beam in vacuum, or in a 2D material. The high SNR and dynamic range are essential for resolving both these strong and weak features, spanning more than 4 orders of magnitude. Because of the detector's high SNR for high-energy electrons and quality of the pedestal subtraction, multiple frames can be summed without a significant impact from systematic noise. This has been demonstrated with integrating pixel array detectors used for x-ray imaging(Philipp *et al.* 2011) and the same principle applies to electron microscopy data. Figures 11c & 10d show the accumulation of data over 10 frames, where the details of the unsaturated central beams and Kikuchi bands are much clearer. Even after summing over 100 frames (Figures 11e,f), there is no noticeable systematic fixed pattern noise. In practice, millions of frames can be summed without significant addition of systematic noise, where the systematic noise in low-fluence (i.e., single-electron) regions in each frame can be suppressed by thresholding without deteriorating the quality of the summed frames. This is an important for imaging radiation-sensitive materials, especially for building up quantitative signals by averaging many low-dose exposures.

Figure 12 shows the high-angle annular dark-field (HAADF), annular bright-field (ABF) and ptychographic phase image of $BaFe_{12}O_{19}$ along the [100] zone axis reconstructed from four-dimensional (4D) datasets recorded with 300 keV electrons and beam current of 15 pA. HAADF and ABF images were synthesized from the same 4D dataset with a focused probe and ptychography used a second dataset with a 20 nm-defocused probe. Both datasets were acquired using a 512×512 scan with a dwell time of 100 µs per pixel, spanning a total acquisition time of 38 seconds. $BaFe_{12}O_{19}$ is a highly insulating material that charges easily under the electron beam, hence the need to keep the beam current low in this instance. Nevertheless, no obvious distortions appear in HAADF (Figure 12a) and ABF (Figure 12b) images, indicating the detector speed outrunning the large sample drift. Multi-slice ptychography, along with position correction algorithms, is used to retrieve the atomic coordinates of both light and heavy elements with high precision(Chen *et al.* 2020, 2021). However, it may not be able to accurately correct large sample drifts using ptychographic algorithms, which can reduce the reconstruction quality. To circumvent the drift issue when using slower detectors, such as the old EMPAD, datasets with a small number of scan points are usually chosen, which limits the field of view. With a faster detector like EMPAD-2G, high-quality ptychographic reconstruction is achieved, shown in Figure 12c, even using a dataset with such a large amount of scan points. In particular, we can identify the Fe-Fe off-mirror-plane displacement with a distance of ~0.35 Å (Cao *et al.* 2015) from the elongated contrast in the ptychographic reconstruction (illustrated as a red elliptical circle on Figure 12c), whereas such structural features cannot be observed in HAADF or ABF images due to the limited resolution. Figure 12c shows a reconstruction using a part of the dataset containing only 64×64 diffraction patterns, but the whole 512×512 scan data is ready for ptychography when the computational resources are available.

As a final example, we show the imaging of order parameters in ferroelectric thin films using the EMPAD-G2. Figure 13 shows the imaging of a $PbTiO_3$ film epitaxially grown on a $DyScO_3$ substrate recorded using a 300 keV electron probe with semi-convergence angle of 2.2 mrad and 2 nA of beam current – a dose rate of 12.5 x $10^9$ e⁻/s. Fig. 13a shows the ADF image of the film reconstructed from the 4D dataset acquired using a 512×512 scan and dwell time of 100 µs per pixel. Inherent from the Poisson statistics, for which the SNR scales with square root of number of electrons recorded in the detector, the large electron beam current was essential for a precise determination of strain and polarization fields. With a maximum of a little over 1.25 million electrons per frame in this experiment, the best achievable precision is about 0.1% of a disk width. It will always be worse than this as the dose is distributed among multiple beams, but it provides a bound and shows the need to record large doses in short time for high-speed mapping. For example, Fig. 13b shows the well-defined HOLZ ring, zero-order Laue zone (ZOLZ) reflections, and the Kikuchi bands all captured simultaneously in 100 µs. The principle of determining polarization using CBED patterns is based on dynamical diffraction effects in which the charge redistribution associated with ferroelectric polarization leads to the intensity asymmetry of Friedel pairs(Deb *et al.* 2020; Zuo & Spence 2017). However, this intensity asymmetry in Bragg reflections may be subject to, or dominated by artifacts such as disinclination strain or crystal mis-tilts, which are inevitable in ferroic perovskites(MacLaren *et al.* 2015; Shao & Zuo 2017). To extract the polarization information (Figs. 13c & 13d), we employ the polarity-sensitive Kikuchi bands which is more robust against crystal mis-tilt artifacts(Shao *et al.* 2021). Simultaneous strain information with a precision of close to 0.1% can be obtained (Fig. 13 e-g) with the exit-wave power Cepstrum (EWPC) analysis of the 4D dataset, which quantitatively measures the changes in projected interatomic distances at each probe position(Harikrishnan *et al.* 2021; Padgett *et al.* 2020).

**Conclusion:**

The EMPAD-G2 allows for rapid acquisition of high dynamic range images, resulting in extremely flexible data analysis, including dark field, bright field, differential phase contrast, and multislice ptychography. The advantages offered by this detector are fast acquisition (i.e., a 10 kHz frame rate), almost no dead times because signal continues to be acquired while the detector is read out, and a high dynamic range even when operated at full speed. These advantages stem from the technology chosen, i.e., direct detection of electrons in a silicon diode coupled to signal processing electronics, and the specific design of the signal processing. One of the design specifics is a charge integrating front-end that allows for high-flux measurements without the drawbacks of counting detectors (e.g., coincidence loss) that put strict limits on the quality of data that can be collected at high speeds with counting detectors. With an integrating front-end, there is no specific signal processing time required for identifying single electron events or losses at high currents. Additionally, the extension of the dynamic range with adaptive gain and incremental (and quantitative) charge removal from the front-end node allows for information to be collected quickly. The importance of these capabilities is clear: high fidelity data from high current probing of a sample can be collected at 10 kHz with minimal deadtime and a high SNR, meaning the impact of sample instabilities is markedly reduced. We have demonstrated strain and polarization mapping at these speeds and introduced an information content metric, the MUIS, that describes the maximum speed a detector can be operated at to obtain a desired SNR. Comparing the MUIS for different design strategies, it becomes clear that pulse counting results in lower frame rates than charge integration for quantitative work that requires high doses, like measuring magnetic and strain fields with high precision. Ultrafast electron diffraction and microscopy, where many electrons can arrive in short bunches, will be another area where this charge integration strategy will be essential for efficient operation. We hope that both this detector and the additional metric to guide the design of future detectors will significantly improve the chances of new scientific observations for electron microscopists.

**Acknowledgements:**

Support for detector development at Cornell includes:

- Thermo Fisher Scientific, Inc.
- The Kavli Foundation
- The W.M. Keck Foundation
- U.S. DOE grants DE-FG02-10ER46693, DE-SC0016035, DE-SC0004079, and DE-SC0017631
- Cornell microscope facilities are supported by U.S. NSF grants DMR-1719875 and DMR-2039380

This project has received funding from the ECSEL Joint Undertaking (JU) under grant agreement No 783247. The JU receives support from the European Union's Horizon 2020 research and innovation program and Netherlands, Belgium, Germany, France, Austria, United Kingdom, Israel, Switzerland.

The oxide test samples were provided by Darrell Schlom and Evan Y. Li, with specimen prep help from Harikrishan K. P. We thank Dr Mariena Sylvestry Ramos for assistance with the Thermo Fisher Scientific Themis, and Bert Freitag for helpful comments. We thank Prafull Purohit for partial schematic entry of digital edge readout structures.

**Competing Interest Statement:**

Funding sources for the detector development, including Thermo Fisher Scientific, are described in the acknowledgements. Thermo Fisher Scientific employees involved in the project are identified by their authorship bylines. We have no other competing interests.

Plotkin-Swing, B., Corbin, G. J., De Carlo, S., … Krivanek, O. L. (2020). Hybrid pixel direct detector for electron energy loss spectroscopy. *Ultramicroscopy*, **217**, 113067.

Shao, Y.-T., Das, S., Hong, Z., … Muller, D. A. (2021). Emergent chirality in a polar meron to skyrmion phase transition. *ArXiv:2101.04545 [Cond-Mat]*. Retrieved from http://arxiv.org/abs/2101.04545

Shao, Y.-T., & Zuo, J.-M. (2017). Lattice-Rotation Vortex at the Charged Monoclinic Domain Boundary in a Relaxor Ferroelectric Crystal. *Physical Review Letters*, **118**(15), 157601.

Tate, M. W., Purohit, P., Chamberlain, D., … Gruner, S. M. (2016). High Dynamic Range Pixel Array Detector for Scanning Transmission Electron Microscopy. *Microscopy and Microanalysis*, **22**(1), 237–249.

Trunk, U., Allahgholi, A., Becker, J., … Zimmer, M. (2017). AGIPD: a multi megapixel, multi megahertz X-ray camera for the European XFEL. In T. G. Etoh & H. Shiraga, eds., *Selected Papers from the 31st International Congress on High-Speed Imaging and Photonics*, SPIE. doi:10.1117/12.2269153

Xu, T., Chen, Z., Zhou, H.-A., … Jiang, W. (2021). Imaging the spin chirality of ferrimagnetic Néel skyrmions stabilized on topological antiferromagnetic Mn3Sn. *Physical Review Materials*, **5**(8), 084406.

Zuo, J. M., & Spence, J. C. H. (2017). *Advanced Transmission Electron Microscopy: Imaging and Diffraction in Nanoscience*, 1st ed. 2017, New York, NY: Springer New York : Imprint: Springer. doi:10.1007/978-1-4939-6607-3

Tables

| | |
|---|---|
| **Pixel Size** | 150 µm × 150 µm |
| **Array Size** | 128 × 128 pixels |
| **Maximum Frame Rate** | 10 kHz |
| **Signal to Noise Ratio (SNR) @ 80 keV, 120 keV, 300 keV for single electron detection** | 31, 46, 115 |
| **Noise in keV equivalent** | 2.6 keV |
| **Acquisition Duty Cycle** | ~ 99% |
| **Maximum current/pixel incident electrons** | >175 pA/pixel at 300 keV <br> >875 pA/pixel at 60 keV |
| **Pixel well depth** | ~ $4 \times 10^8$ keV <br> ≈ $1.4 \times 10^6$ 300 keV $e^-$ |
| **Dynamic Range** | $1.3 \times 10^7$ at 10 kHz frame rate |

**Table 1**

Figures

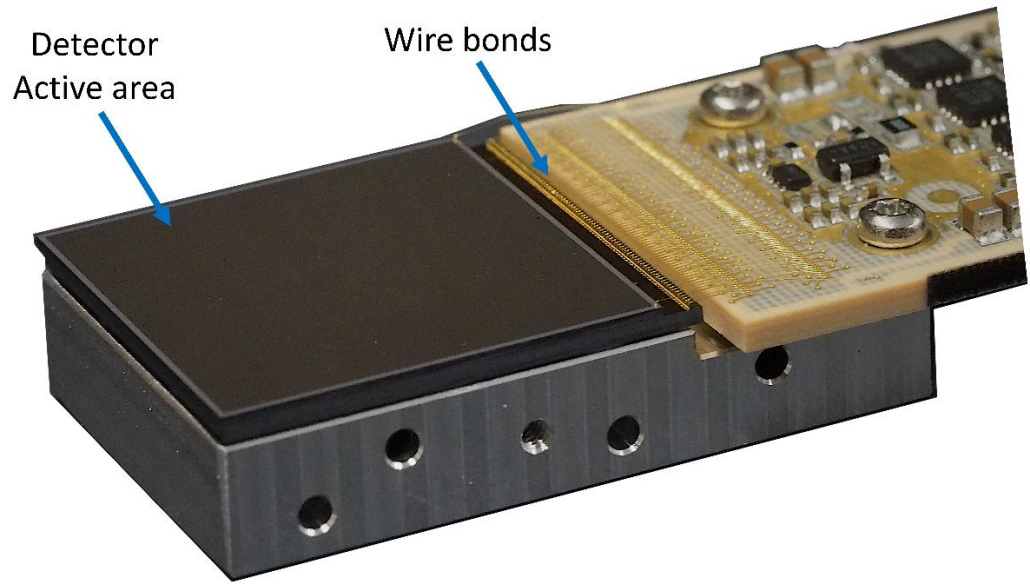

**Figure 1:** Picture of the detector module showing the active imaging area and wire bonds along one edge. The single-edge connection simplifies future tiled detector designs.

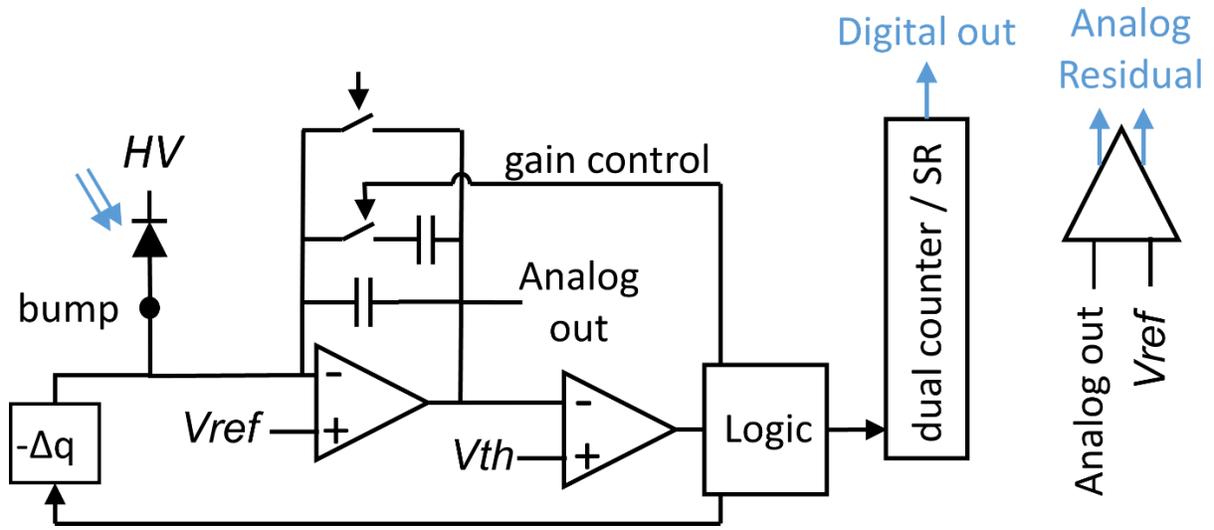

**Figure 2:** High-level pixel diagram. The bump bond and sensor diode are shown schematically on the left. The charge integrating front-end actively adapts during integration if a threshold is crossed, first reducing the gain by adding a feedback capacitor, then removing charge in fixed increments.

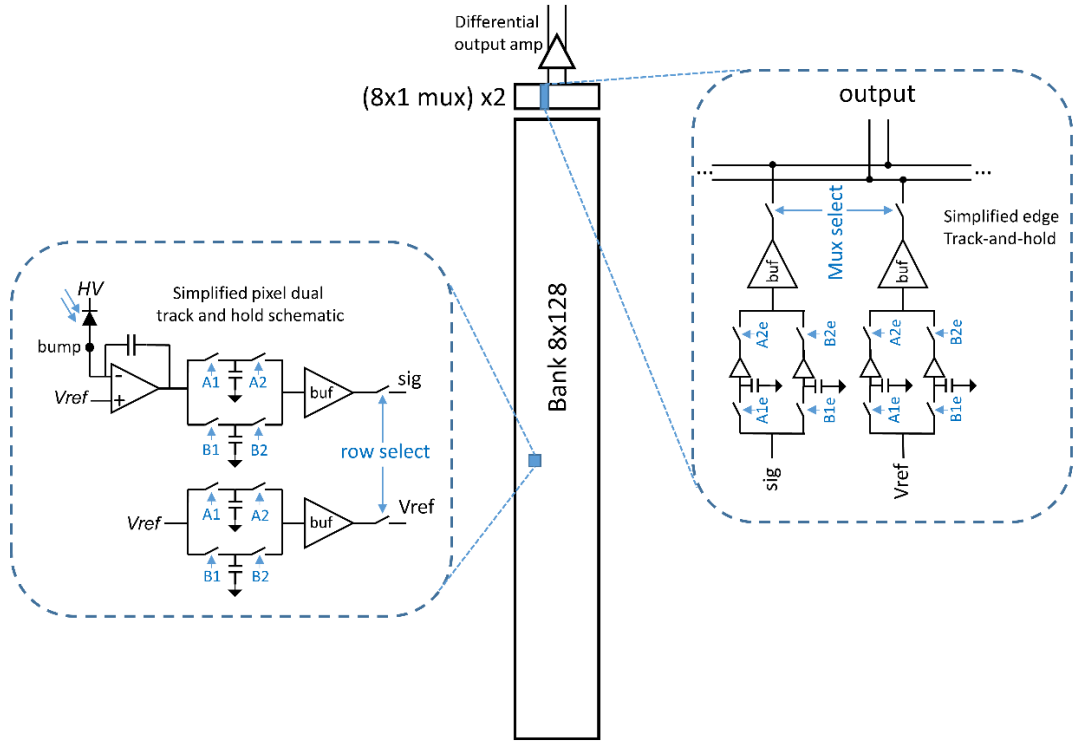

**Figure 3:** A high-level schematic analog output – column level

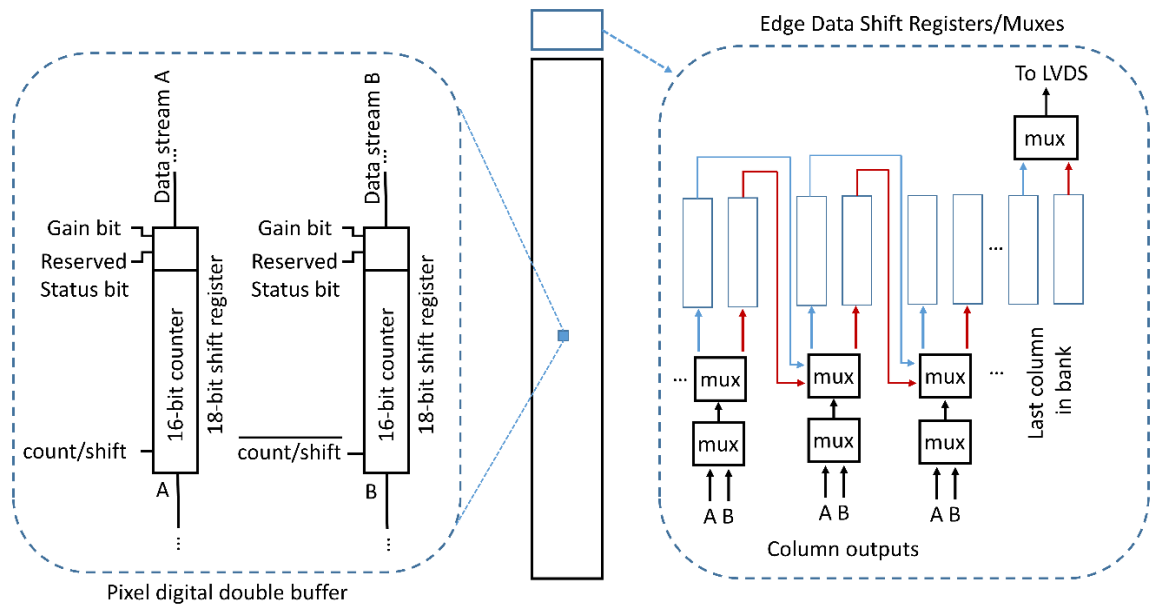

**Figure 4:** A high-level schematic digital readout – column level

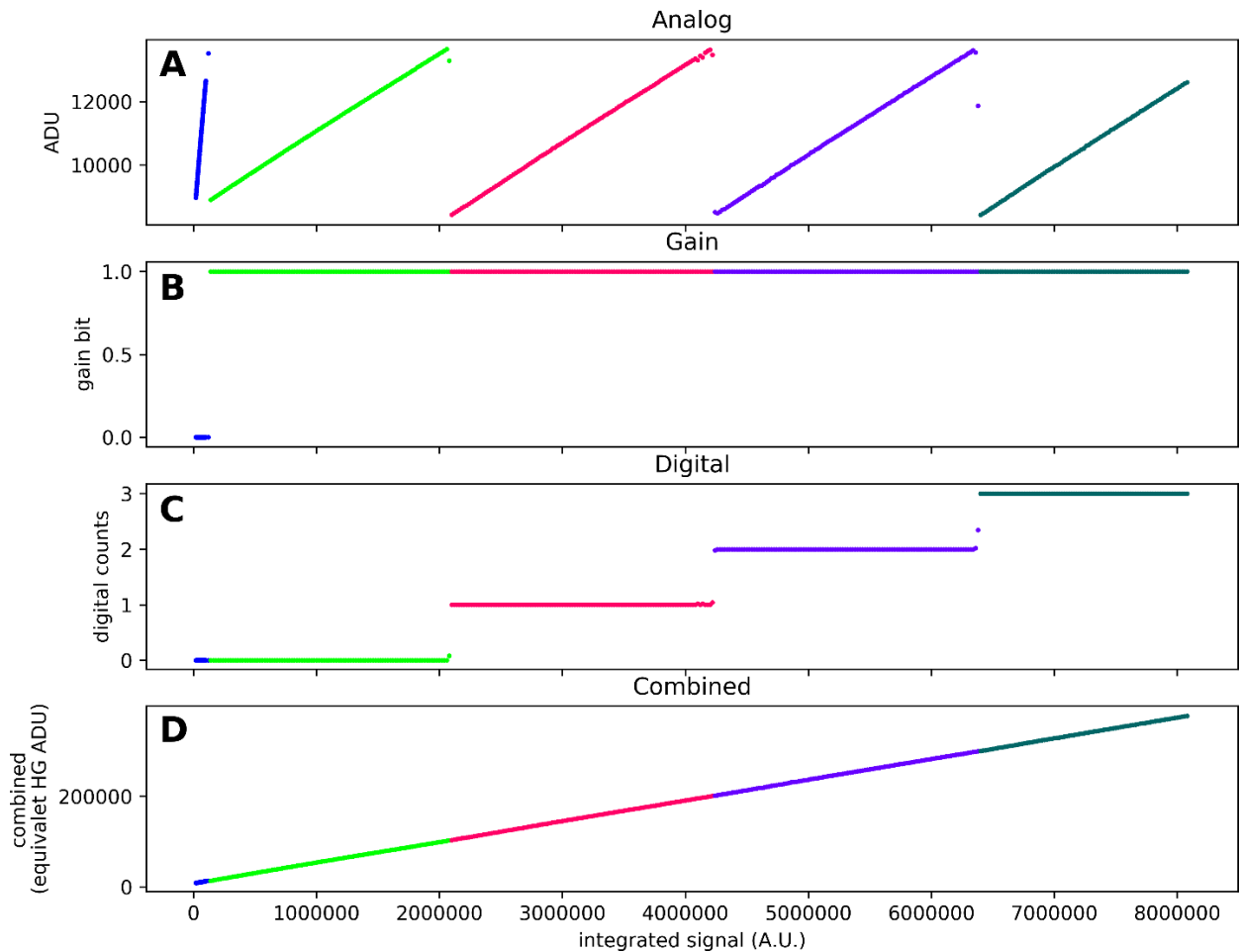

**Figure 5:** Data recorded in a single pixel under uniform illumination with increasing integration time. The raw data includes an analog value (A) from the voltage remaining on the integration capacitor, a gain bit (B), and a 16-bit digital number (C) corresponding to the number of charge dump cycles. These data are scaled together (D) to obtain calibration constants which provide a continuous linear measurement of charge carriers produced in the sensor diode. The data shown were obtained using an optical LED array (Bridgelux, BXRC-50C1001-D-74-SE), run at a constant current, providing an optical flood field. Each point is an average of 50 readings and accounts for non-digital integer counts. The blue trace is acquired in high gain, before the pixel is triggered to flip the gain bit. The light green trace is after the pixel has adaptively switched to a lower gain. The red trace is the first digital count. Colors to the right of the red traces represent subsequent increments of the pixel's digital counter.

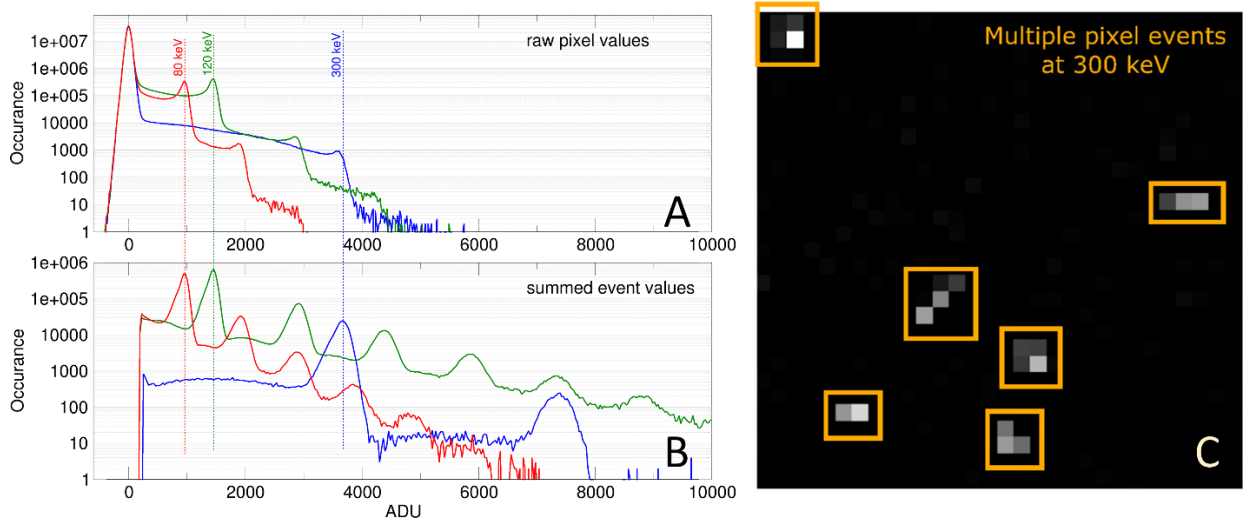

**Figure 6:** Low-fluence histograms. (A) histograms of raw pixel values across the array for three incident electron energies (80 keV, 120 keV, 300 keV) showing the zero-signal peak and discrete energy peaks. (B) histograms of signal from identified connected clusters of pixels that detected electron events. Histograms of connected pixels that recombine charge that was split between pixels. (C) a sub-section of a single low-fluence image of 300 keV electrons, showing the splitting of charge between multiple pixels.

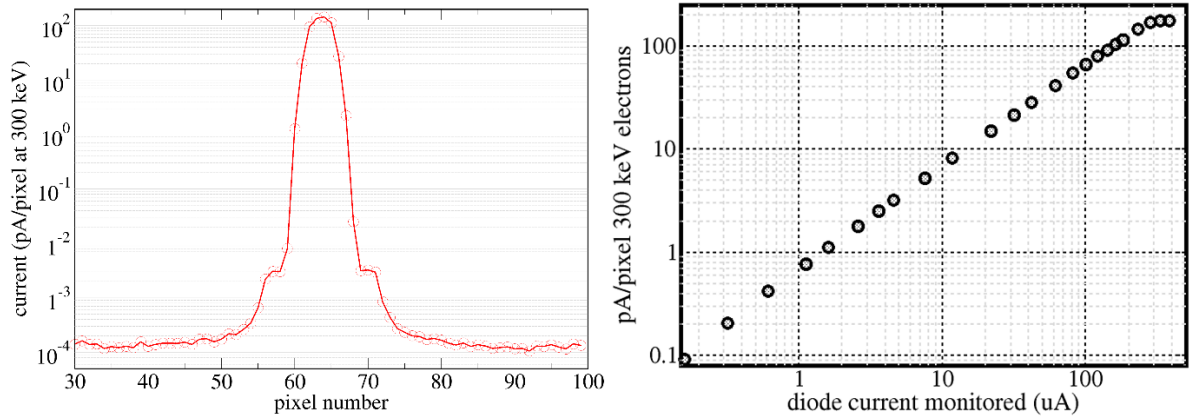

**Figure 7:** (left) Cross section of a small focused spot of 300 keV electrons used in linearity measurements. Data taken from the average of 1000 images with 100 us exposure time at 182 uA sensor current and a maximum incident 300 keV electron current of 144 pA/pixel. Intensity in the peak was > 60,000 electrons/pixel/frame, whereas the tail regions had < 0.03 electrons/pixel/frame on average. (right) Intensity in the brightest pixel as a function of varying beam current. Intensity has been converted to incident electron current per pixel. The pixel shows linear response up to 175 pA/pixel of incident beam current at 300 keV.

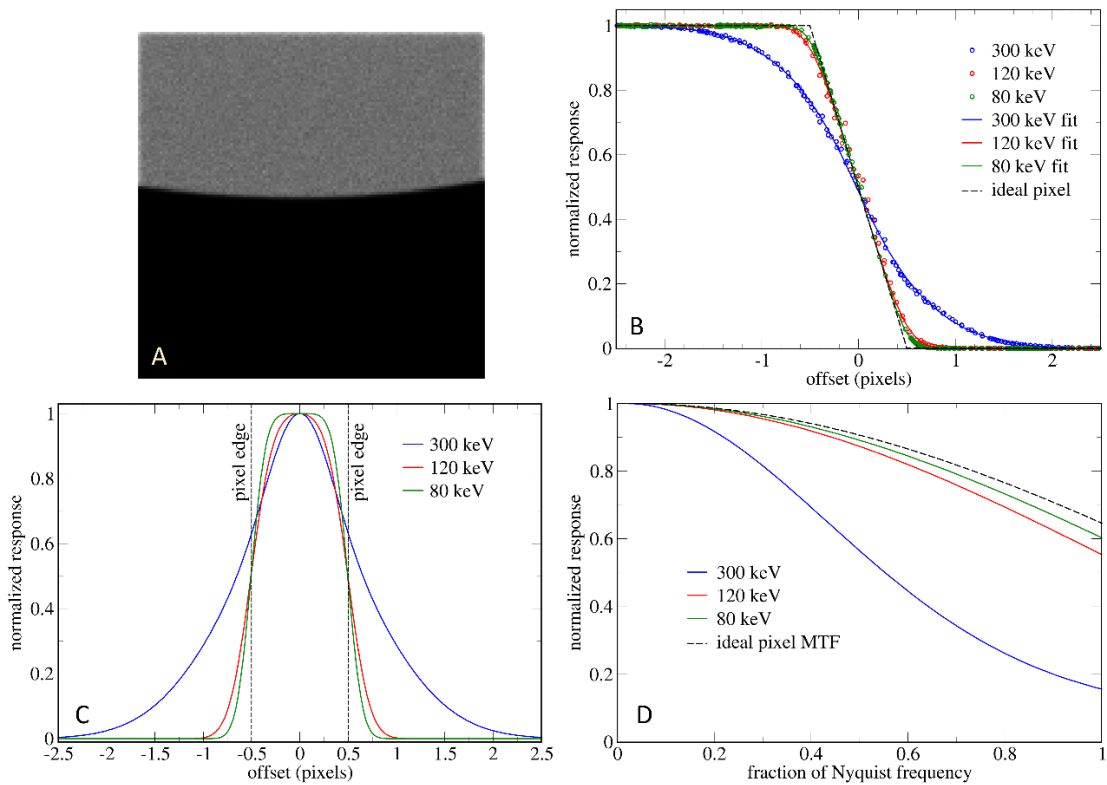

**Figure 8:** Measurements of the spatial response at the different energies. (upper left, A) an image of an aperture projected onto the detector face. (upper right, B) The edge response the detector at 300 keV, 120 keV, and 80 keV, as a function of the distance from the imaged aperture edge. The best fit circular arc was used to define the edge of the aperture. The ideal pixel response corresponds to an ideal pixel, i.e., no charge spread. (lower left, C) The line spread function (LSF) at 300 keV, 120 keV, and 80 keV found by differentiating the fitted function to the edge response. (lower right, D) The modulation transfer function (MTF), derived from the FFT of the LSF, plotted up to the Nyquist frequency.

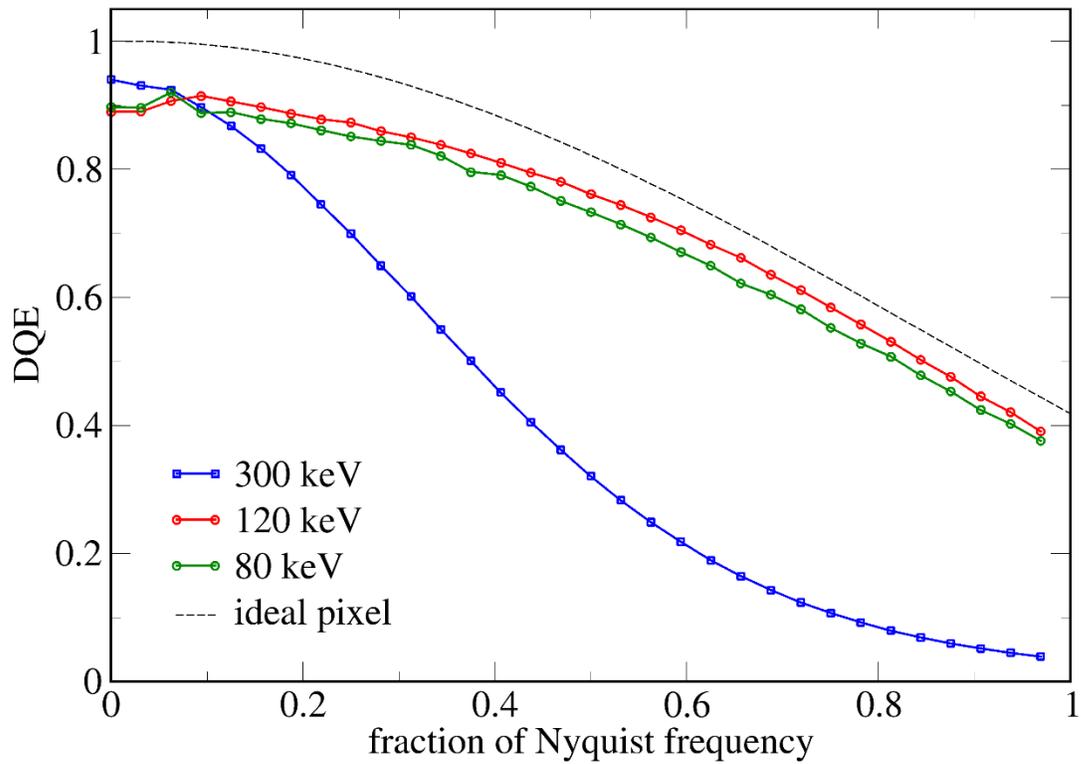

**Figure 9:** Figure 9: DQE(ω) for 300 keV, 120 keV and 80 keV. Also shown is the ideal response for a detector with 150 µm pixilation. The DQE(0) for the 300 keV, 120 keV, and 80 keV are 0.94, 0.90, and 0.90, respectively.

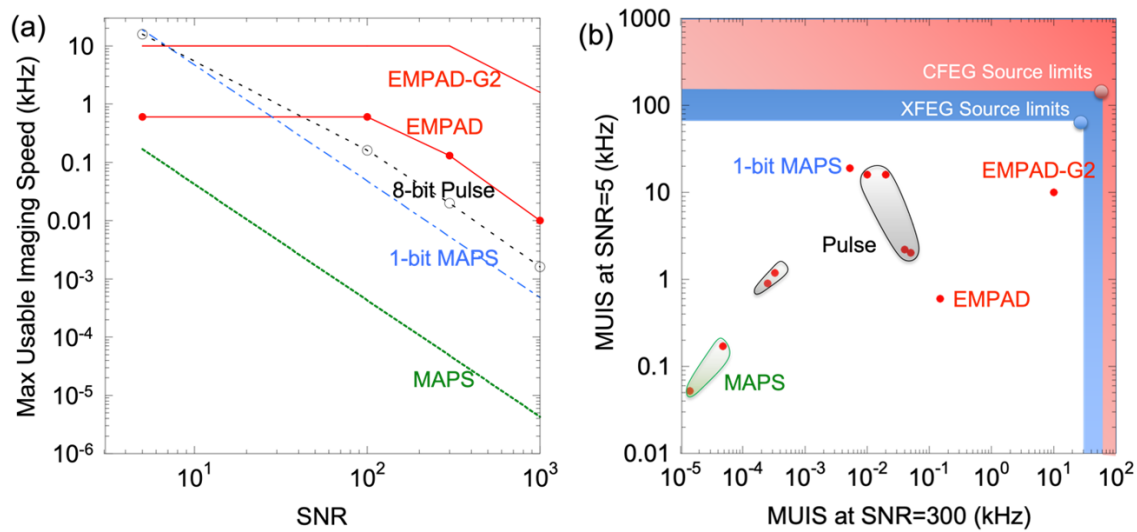

**Figure 10**: Maximum Usable Imaging Speed (MUIS): (**a**) MUIS as function of SNR for the different detector strategies discussed in the text. 8-bit pulse models a modern pulse counting PAD, MAPS models a large-pixel format MAPS detector optimized for cryoelectron microscopy and 1-bit MAPS models a high-speed detector designed to operate at an 87 kHz frame rate but with only a 1-bit readout. (**b**) A summary of achievable MUIS for different detector designs, represented by a high-speed but noisy imaging mode at SNR=5 in all pixels intended to reflect TEM usage or basic atomic-resolution STEM imaging, and a quantitative mapping mode at SNR=300 in a few pixels (see text) such as for strain and diffraction measurements. Here, a range of dead-times, readout and data-transfer schemes are explored for the different pulse counting options. Deliverable beam currents limits set by the source performance are also shown for thermal field and cold field emitters. Currently system performance is still limited by the detectors and not the sources. In both panels, these are not a direct pixel-to-pixel comparisons, but instead assumes that the other designs, which typically have smaller pixels, have been binned down to match the EMPAD pixel count. Without re-binning, their MUIS performance would drop 4-16 x compared to what is plotted here.

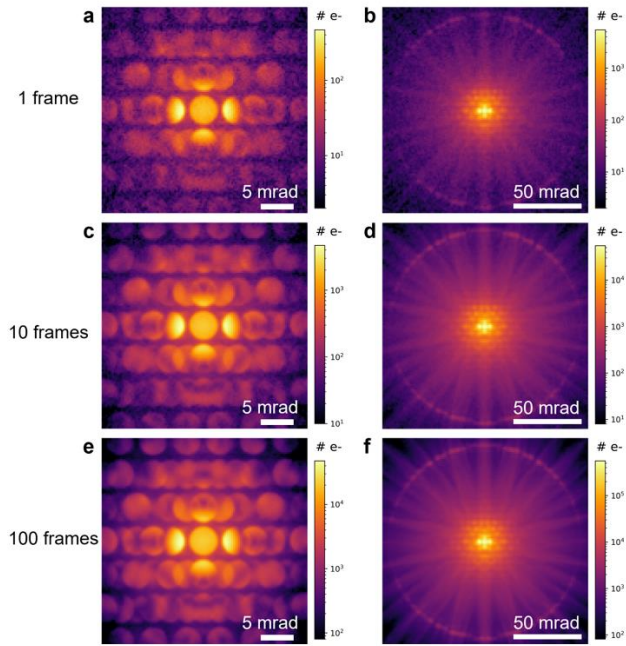

**Figure 11**: High-current, high-dynamic range imaging: CBED patterns of TbScO$_3$ recorded using the EMPAD-G2 at 300 keV with 1 nA of beam current, 100 µs dwell time and summed over (**a, b**) 1 frame, (**c, d**) 10 frames and (**e, f**) 100 frames for two different camera lengths, and total acquisition times of 0.1, 1, 10 ms respectively. All CBED images show the number of electrons detected on the G2-EMPAD, showing quantitative electron counting.

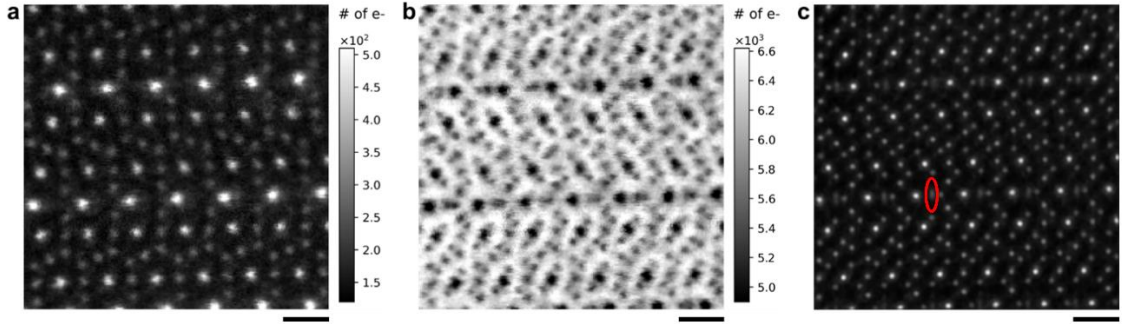

**Figure 12**. Comparing atomic resolution images of $BaFe_{12}O_{19}$. (a) Atomic resolution high-angle annular dark-field (HAADF) and (b) annular bright-field (ABF) images of highly-insulating $BaFe_{12}O_{19}$ acquired using the EMPAD-G2 with 300 keV electron beam and current of 15 pA to reduce sample charging. The four-dimensional dataset was acquired using a 512×512 scan with a 100 µs dwell time per pixel, spanning a total acquisition time of 38 seconds. (c) Phase image from the multi-slice ptychography reconstruction from the 4D dataset using a defocused probe. The sample thickness is about 14 nm estimated from ptychography. To show the same field of view as the ptychography, (a, b) were cropped to 210×210 pixels from the 512×512 scan – an equivalent of 4.5s acquisition time. Scale bars, 5 Å. The red elliptical circle illustrates the Fe-Fe off-mirror-plane displacement.

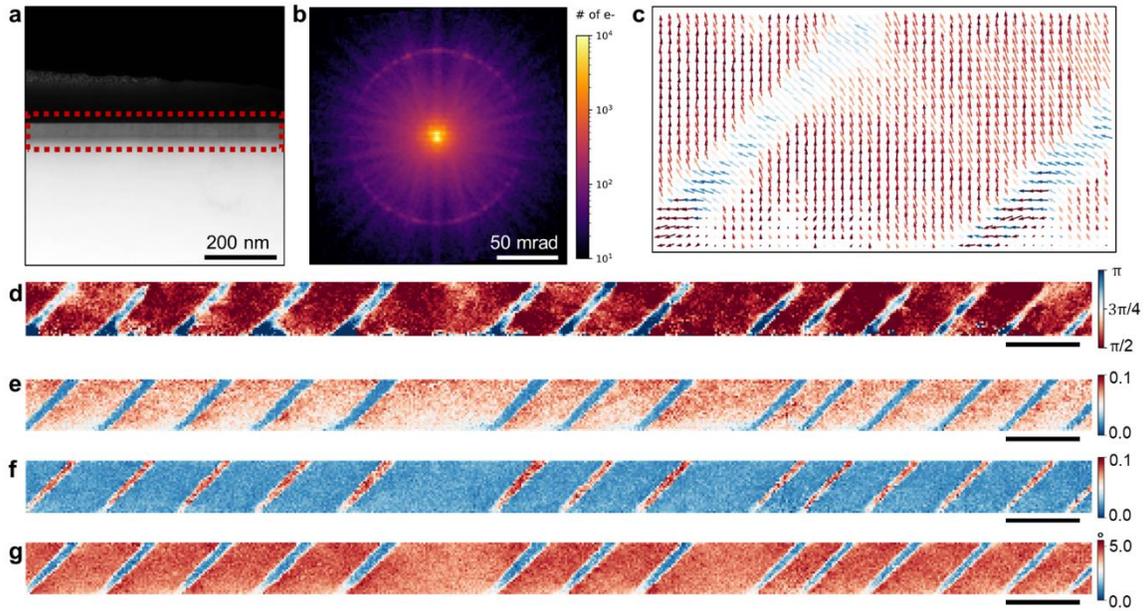

**Figure 13:** Large field of view, high-speed simultaneous mapping of ferroelectric polarization and strain fields in an epitaxial PbTiO$_3$ film grown on a DyScO$_3$ substrate. The four-dimensional dataset is acquired using the EMPAD-G2 with a 512×512 scan, a 100 µs dwell time per pixel, and a large beam current of 2 nA. (a) HAADF-STEM image reconstructed from the 4D-dataset. (b) Experimental diffraction pattern of DyScO$_3$ [101] recorded within 100 µs showing the HOLZ ring, Kikuchi bands and unsaturated central beam. (c, d) Polarization maps extracted from the intensity asymmetry in the Kikuchi bands of PTO layer. For clarity, Figure (c) shows the enlarged region of 50×26 pixels (0.13 seconds total dwell time), while (d) shows the 512×26 pixels (1.33 seconds total dwell time) for the polarization map. Strain maps obtained from the same dataset by employing Cepstral transform strain analysis, showing (e) $\varepsilon_{11}$, (f) $\varepsilon_{22}$, and (g) in-plane rotation θ, respectively. Scale bars in (e-g): 50 nm.